\documentstyle[twoside,fleqn,espcrc2,epsf]{article}
\title{Light Spectrum and Decay Constants in Full QCD with 
       Wilson Fermions\thanks{talk presented by U.~Gl\"assner}}
\author{SESAM-Collaboration: U.~Gl\"assner$^{\rm a}$,
        S.~G\"usken$^{\rm a}$, H.~Hoeber$^{\rm b}$, Th.~Lippert$^{\rm b}$, 
        G.~Ritzenh\"ofer$^{\rm b}$, K.~Schilling$^{\rm a,b}$, 
        G.~Siegert$^{\rm b}$ and A.~Spitz$^{\rm a}$. \\[8pt]
{\rm $^a$}Physics Department, University of Wuppertal, D-42097
           Wuppertal, Germany,\\[8pt]       
{\rm $^b$}HLRZ c/o Forschungszentrum J\"ulich, D-52425 J\"ulich,
          and DESY, D-22603 Hamburg, Germany.}  
\begin{document}
\begin{abstract} 
We present results from an analysis of the light spectrum and the
decay constants $f_{\pi}$ and $f_V^{-1}$ in Full QCD with $n_f=2$ Wilson
fermions at a coupling of $\beta = 5.6$ on a $16^3 \times 32$ lattice.
\end{abstract} 
\maketitle

\section{INTRODUCTION}

This analysis was performed as part of the SESAM project
\cite{melbourne} to investigate {\bf S}ea quark {\bf E}ffects on {\bf
  S}pectrum {\bf A}nd {\bf M}atrix-elements.  Configurations were
generated using the Hybrid-Monte-Carlo algorithm with the standard
Wilson action at a coupling of $\beta=5.6$ on a $16^3 \times 32$
lattice. We work at three different values of the dynamical quark
mass: $\kappa_{\rm{sea}}=0.1560$, $0.1570$ and $0.1575$ corresponding
to $m_{\pi}/m_{\rho}$-ratios of $0.83(1)$, $0.76(1)$ and $0.71(2)$
\cite{spectrum}.  Our lightest quark-mass is approximately equal to
the strange quark-mass.  Up to now we generated about 8000 trajectories
of unit-length with a time-step of $\Delta t=0.01$. After measuring
the integrated autocorrelation time for the plaquette and the
pion-correlator at fixed timeslice \cite{spectrum,potential} we
decided to use configurations separated by 25 trajectories for the
present spectrum analysis. Our sample consists of $100$, $160$ and
$100$ configurations for our 3 sea-quark values. Since we aim at a
final sample of $3 \times 200$ configurations this talk should be
considered as a half-time status report.

Quark propagators for the set of valence $\kappa$-values \{ 0.1555,
0.1560, 0.1565, 0.1570, 0.1575\} were computed using the standard
overrelaxed Minimal Residual Algorithm with smeared sources and local
as well as smeared sinks. We used the Wuppertal gauge-invariant
gaussian smearing method with $N=50$ iterations and $\alpha=4$,
fitting the hadron-correlators to single-exponential functions. 
Throughout this analyis we neglect correlations in time and in
$\kappa$. We hope to present a stable correlated analysis on our
final sample. 

\section{CHIRAL EXTRAPOLATIONS WITH FIXED SEA-QUARK MASS} 

\begin{table*}
\begin{center}
  \begin{tabular}{llllllll}
  $\kappa_{\rm{sea}}$ & $\kappa_C$ & $a M_\rho$ & $a_{\rho}^{-1}$ [GeV] & 
  $M_N$ [GeV] & $M_{\Delta}$ [GeV] & $f_{\pi}$ [MeV] & $f_V^{-1}$ \\ 
  \hline
  0.1560 & 0.16065(8) & 0.359(8) & 2.14(5)  & 1.09(7) & 1.30(9) 
         & 125(9)  & 0.33(3)\\
  0.1570 & 0.15987(6) & 0.341(8) & 2.23(7)  & 1.14(6) & 1.40(10) 
         & 101(8)  & 0.35(2)\\
  0.1575 & 0.15963(11) & 0.316(10) & 2.44(8) & 1.09(10) & 1.22(8) 
         & 118(15) & 0.30(3)
  \end{tabular}
  \caption{\it Extrapolation results with fixed sea-quark mass.
           Errors quoted are purely statistical\label{results}}
\end{center}
\end{table*}
To determine the critical hopping-parameter $\kappa_c$ at a given
value of $\kappa_{\rm{sea}}$ we fit the pseudoscalar masses according
to
\begin{equation} \label{pion}
M_{PS}^2 = b m_q + c m_q^2\\ 
\end{equation}
with $m_q=1/2(1/\kappa - 1/\kappa_c)$ being proportional to the quark
mass. We find $c$ to be significantly different from zero: a linear
fit leads to a decrease of $\kappa_c$ by four standard deviations with
respect to the quadratic fit. The resulting values of $\kappa_c$ are
shown in table \ref{results} together with their statistical errors.
To estimate systematic errors on $\kappa_c$, we exclude the largest
quark-mass from our data and find a change in $\kappa_c$ of the order
of one standard deviation.

\begin{figure}
  \epsfxsize=7.5cm\epsfbox{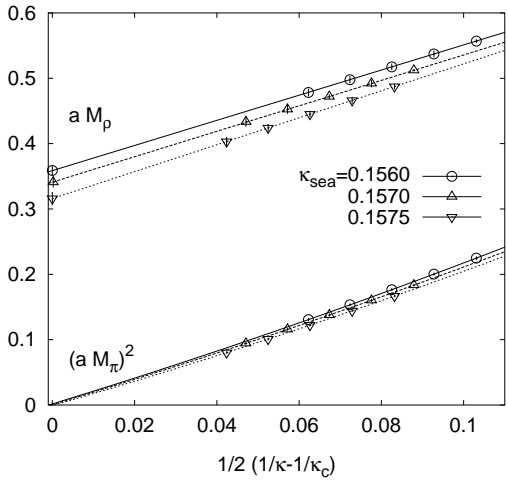}
  \caption{\it Meson extrapolations with fixed sea-quark 
           mass.\label{mesons}}  
\end{figure}

For the chiral extrapolation of vector masses we use two different
parametrizations
\begin{eqnarray} 
M_{V} & = & m_{\rho} + b m_q + c m_q^2 \label{rho_sq}\\
M_{V} & = & m_{\rho} + b m_q + c m_q^{3/2}\label{rho_32} ,
\end{eqnarray}
which are motivated by chiral perturbation theory.  In contrast to the
$\Pi$ extrapolation we find that the values obtained from 2- and
3-parameter fits agree within errors; moreover eqs. \ref{rho_sq} and
\ref{rho_32} yield identical results. In table \ref{results} we quote
the values for the $\rho$-masses and the corresponding
lattice-spacings as obtained from the 2-parameter-fits.  The masses
are calculated at $\kappa_c$ because we find the mass-shifts induced
by $\Delta \kappa = \kappa_c - \kappa_{\rm{light}}$ to be negligible
compared to the statistical errors\footnote{In fact $\kappa_c$ and
  $\kappa_{\rm{light}}$ still agree within errors}. We will discuss
the light quark masses estimated from $\Delta \kappa$ in detail in a
fourthcoming publication \cite{spectrum}.  Figure \ref{mesons}
illustrates the variation of $m_{\rho}$ with $m_q$ and
$\kappa_{\rm{sea}}$.

\begin{figure}
  \epsfxsize=7.5cm\epsfbox{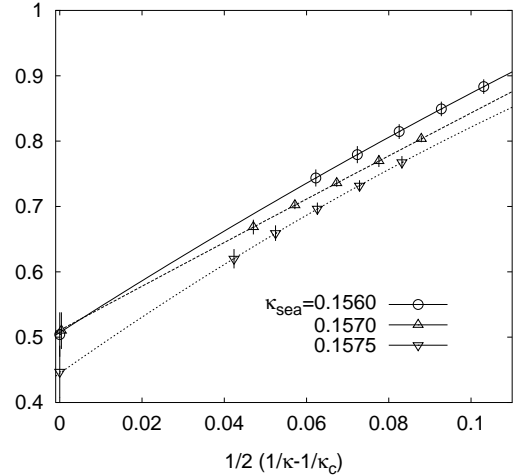}
  \caption{\it Nucleon extrapolations with fixed sea-quark 
           mass.\label{nucleon}}
\end{figure}

The nucleon data (see figure \ref{nucleon}) cannot be fitted with
$c=0$, but ans\"atze eqs. \ref{rho_sq} and \ref{rho_32} do equally
well. The quadratic extrapolation to $\kappa_c$ leads to the values
$M_N$ quoted in table \ref{results} together with the values for
$M_{\Delta}$. With the present statistics we do not see any sea-quark
dependence of the nucleon mass in physical units.

We now consider the decay-constants, $f_{\pi}$ and $f_V^{-1}$, 
defined as 
\begin{eqnarray}
<0|A|\pi> Z_k Z_A &=& f_{\pi} m_{\pi} \label{fpi}\\
<0|V_i|\rho> Z_k Z_V &=&  \epsilon_i f_V^{-1} m_{\rho}^2  \label{fv}.
\end{eqnarray}
$A$ and $V_i$ are the local axial and vector currents on the lattice
and $Z_k$, $Z_A$ and $Z_V$ the renormalization constants which connect
the expectation values of equation \ref{fpi} and \ref{fv} to the
continuum. We determine these renormalization constants using
tadpole improved perturbation theory \cite{spectrum,lepage}.
\begin{figure}
  \epsfxsize=7.5cm\epsfbox{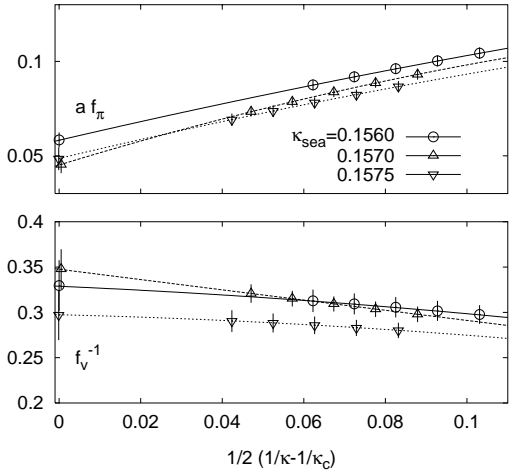}
  \caption{\it Pseudoscalar and Vector decay constants with 
           fixed sea-quark mass. \label{decay}}
\end{figure}
The $m_q$-dependence of the decay-constants is presented in figure
\ref{decay}. Obviously the statistical accuracy is not yet high enough
to resolve a $\kappa_{\rm{sea}}$-dependence of $f_{\pi}$ and $f_V^{-1}$
at the chiral point (see also table \ref{results}). 

\section{EXTRAPOLATIONS IN {\bf \large $\kappa_{\rm{sea}}$}}

\begin{figure}
  \epsfxsize=7.5cm\epsfbox{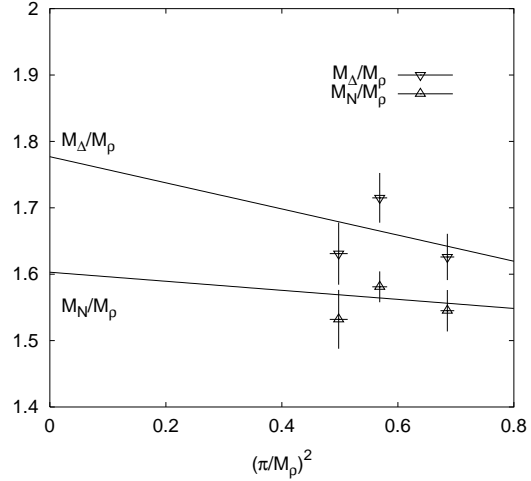}
  \caption{\it Extrapolations of mass ratios with 
           $\kappa_{\rm{sea}}=\kappa_{\rm{val}}$.\label{sim}}
\end{figure}

A consistent method to extrapolate to zero {\it sea}-quark mass is to
use only the data-points with $\kappa_{\rm{sea}}=\kappa_{\rm{val}}$
for the extrapolation. In the case of dimensionful observables one
might argue about the impact of the varying scale along the
trajectory. For this reason we consider only mass-ratios as shown in
figure \ref{sim}. We use linear fits only. The extrapolated values,
quoted in table \ref{simtable}, agree (within large errorbars) with
those obtained from extrapolations at fixed $\kappa_{\rm {sea}}$.
\begin{table}
\begin{center}
\begin{tabular}{l|l}
\hline
$M_N/M_{\rho}$        & 1.6(2) \\
$M_{\Delta}/M_{\rho}$ & 1.8(2) \\
$f_{\pi}/M_{\rho}$    & 0.14(3)\\ 
$f_V^{-1}$            & 0.32(6)\\
\hline
\end{tabular}
  \caption{\it Results from extrapolations in $\kappa_{\rm{sea}}$.
          \label{simtable}}
\end{center}
\end{table}

\section{CONCLUSIONS AND OUTLOOK}

On our present statistics, uncorrelated data analyses of $t$- and
$\kappa_{val}$ distributions allow for stable chiral extrapolations in
valence quark mass, $m_q$.  We find the $\rho$-meson mass to be be
consistent with a linear chiral ansatz, while the remaining quantities
considered exhibit {\it nonlinear} behaviour in $m_q$. $m_{\rho}$
varies on the level of 15 \% across our range of sea quark masses.
Baryon masses and decay constants so far indicate a trend but produce
no compelling evidence for such variation.

Chiral extrapolations of our present data in $\kappa_{\rm{sea}}$ itself
still carry too large uncertainties to be reliable. Better statistics
and deeper penetration into the chiral regime are needed to improve on
this situation.


\begin{thebibliography}{99}
\bibitem{melbourne}
SESAM-Collaboration, Nucl.~Phys.~B (Proc. Suppl.) {\bf 47} (1996) 386.
\bibitem{spectrum}
SESAM-Collaboration, in preparation
\bibitem{potential}
SESAM-Collaboration, Preprint HLRZ 96-20 and WUB 96-17, to be
published in Phys.~Lett.~B .
\bibitem{lepage}
G.P.~Lepage and P.B.~Mackenzie, Phys.~Rev.~D {\bf 48} (1993) 2250.
\end{thebibliography}
\end{document}